\documentclass[twocolumn,showpacs,fleqn,nobibnotes]{revtex4}

\usepackage{amsmath}
\usepackage{graphicx}
\usepackage{float}
\usepackage{subfigure}

\def\lsim{\raise0.3ex\hbox{$<$\kern-0.75em\raise-1.1ex\hbox{$\sim$}}}
\def\gsim{\raise0.3ex\hbox{$>$\kern-0.75em\raise-1.1ex\hbox{$\sim$}}}

\newcommand{\rr}{\mbox{\boldmath $r$}}

\begin{document}

\title{Exclusive photoproduction of $J/\psi$ and $\psi(2S)$ states in proton-proton collisions at the CERN LHC}
\author{M.B. Gay Ducati, M.T. Griep and M.V.T. Machado}

\affiliation{High Energy Physics Phenomenology Group, GFPAE  IF-UFRGS \\
Caixa Postal 15051, CEP 91501-970, Porto Alegre, RS, Brazil}

\begin{abstract}
In this work we investigate the exclusive photoproduction of $J/\psi$ and the radially excited $\psi (2S)$ state off nucleon in proton-proton collisions.  The theoretical framework considered in the analysis is the light-cone dipole formalism and predictions are done for proton-proton collisions at the CERN-LHC energy of 7 TeV. The theoretical uncertainties are investigated and comparison is done to the recent LHCb  Collaboration data for the exclusive charmonium production.

\end{abstract}

\maketitle

\section{Introduction}

The exclusive vector meson photoproduction, $\gamma +p\rightarrow V+p$, has being investigated both experimentally and theoretically in recent years as it allows to test perturbative Quantum Chromodynamics. The quarkonium masses, $m_V$, give a perturbative scale for the problem even in the photoproduction limit, $Q^2=0$.  An important feature of these processes at the high energy regime is the possibility to investigate the Pomeron exchange. For this energy domain hadrons and photons can be considered as color dipoles in the mixed light cone representation \cite{nik}, where their transverse size can be considered frozen during the interaction. Therefore, the scattering process is characterized by the color dipole cross section describing the interaction of those color dipoles with the nucleon target. Dipole sizes of transverse size $r\approx 1/\sqrt{m_V^2+Q^2}$ are probed by the $1S$ vector meson production amplitude \cite{nik} whereas the  diffractive production of the $2S$ radially excited vector mesons present the so-called node effect \cite{Nemchik:1996pp}. A strong cancellation of dipole size contributions to the production amplitude from the region above and below the node position in the $2S$ radial wavefunction \cite{Nemchik:2000de} takes place and gives origin to a large suppression of the photoproduction  $2S$ states compared to the $1S$ states.

In the present work we focus on the exclusive photoproduction of $J/\psi$ and the radially excited $\psi (2S)$ mesons off nucleon in proton-proton collisions. An important motivation is the recent measurement by the LHCb Collaboration of the cross section at $\sqrt{s}=7$ TeV of the exclusive dimuon final states, including the $\psi (2S)$ state \cite{LHCb}. Those measurements were performed at forward rapidities $2.0\leq \eta_{\mu^{\pm}}\leq 4.5$, which corresponds to sufficiently Bjorken-$x$ variable down to $x\approx 5\times 10^{-6}$. The theoretical framework considered in the current investigation  is the light-cone dipole formalism, where the $c\bar{c}$ fluctuation (color dipole) of the incoming quasi-real photon interacts with the nucleon target via the dipole cross section and the result is projected in the wavefunction of the observed hadron. At high energies, a transition of the regime described by the linear dynamics of emissions chain to a new regime where the physical process of recombination of partons becomes important is expected. Such energy regime is characterized by the limitation on the maximum phase-space parton density that can be reached in the hadron wavefunction: the  parton saturation phenomenon. The transition is set by saturation scale $Q_{\mathrm{sat}}\propto x^{\lambda}$. For recent reviews on these subjects we quote Refs. \cite{hdqcd}. Therefore, the theoretical investigation of $\psi(1S)$ and $\psi (2S)$ mesons can shed light on the experimental constraints for the dipole-proton cross section and on the phenomenological models based on parton saturation ideas. Along these lines, recently in Ref. \cite{GGM} we have investigated the photoproduction of radially excited vector mesons off nuclei in heavy ion relativistic collisions. There, the exclusive photoproduction of $\psi(2S)$  off nuclei was analyzed, evaluating the coherent and the incoherent contributions to that process. The approach gives a reasonable description of ALICE Collaboration data for \cite{ALICE1,ALICE2} $J/\psi$ production at 2.76 TeV in PbPb collisions and predictions are provided for the $\psi (2S)$ state.

The aim of this work is twofold. First, we show predictions for the photoproduction of $\psi(1S)$ and its excited state in proton-proton collisions at the LHC within the same phenomenological formalism. This fact is completely new as most part of predictions in literature concern only to the Psi(1S) state. Second, we investigate the sensitivity of recent LHCb data to the small-x dynamics encoded in our phenomenological model for the dipole cross section. As already mentioned the $\psi(2S)$ wavefunction presents nodes compared to the 1S state. This fact turns out the production amplitude sensitive to large dipole size configurations and therefore it is scanning the transition region between the color transparency behavior, $\sigma_{dip} \propto r^2$, to the soft non-perturbative region. In the saturation models such a transition is driven by the saturation scale and this is an clear advantage of phenomenological model considered here. The analysis presentation is organized as follows. In the next section we present a brief review of the diffractive photoproduction of vector mesons in proton-proton collisions. In Section \ref{resultados} we present the predictions for the $\psi(1S)$ and $\psi(2S)$ photoproduction cross sections at forward rapidities.  We compare the theoretical results to the recent LHCb Collaboration  measurements of exclusive $J/\psi$ and $\psi(2S)$ production \cite{LHCb}. In addition, we compare the current results to related approaches available in the literature and discuss the main theoretical uncertainties. Finally, in Section \ref{conc} we summarize our main results and conclusions.

\section{Exclusive meson photoproduction in proton-proton collisions}
\label{coerente}

The exclusive vector meson photoproduction is described by the photon-Pomeron process, $\gamma +p \,(\rightarrow \gamma + I\!P) \rightarrow V+p$. The corresponding production cross section at photon level is denoted by $\sigma (\gamma p\rightarrow V+p)$. Accordingly, the cross section for the exclusive meson photoproduction in hadron-hadron collisions can be factorized in terms of the equivalent flux of photons of the hadron projectile and photon-target production cross section \cite{upcs}. The photon energy spectrum, $dN_{\gamma}/d\omega$, is given by a modified version of Wiezs\"{a}cker-Williams approximation \cite{upcs}
\begin{eqnarray}
\frac{dN_{\gamma}(\omega)}{d\omega}  & =  & \frac{\alpha_{em}}{2\pi\,\omega}\, \left[1+\left(1-\frac{2\omega}{\sqrt{s}} \right)^2\right] \nonumber \\
& \times & \left(\ln \xi - \frac{11}{6} + \frac{3}{\xi}-\frac{3}{2\xi^2}+\frac{1}{3\xi^3}\right),
\label{fluxint}
\end{eqnarray}
where  $\omega$ is the photon energy and $\sqrt{s}$ is the hadron-hadron centre-of-mass energy. Given the Lorentz factor of a single beam, $\gamma_L=\sqrt{s}/(2m_p)$, one has that $\xi = 1+ (Q_0^2/Q_{\mathrm{min}}^2)$ with $Q_0^2=0.71$ GeV$^2$ and $Q_{\mathrm{min}}^2=\omega^2/\gamma_L^2$.

The rapidity distribution $y$ for charmonium  photoproduction, i.e. the $\psi(1S)$ and $\psi (2S)$ states, in  proton-proton collisions can be  written down as,
\begin{eqnarray}
\frac{d\sigma}{dy}(pp \rightarrow   p\otimes \psi \otimes p) & = & S_{\text{gap}}^2 \left[\omega \frac{dN_{\gamma}}{d\omega }\sigma (\gamma p \rightarrow \psi(nS) +p)\right.  \nonumber \\
 &+& \left.  \left(y\rightarrow -y \right) \right], 
\label{dsigdy}
\end{eqnarray}
where $\otimes$ represents the presence of a rapidity gap. The produced state with mass $m_V$ has rapidity $y\simeq \ln (2\omega/m_V)$ and the square of the $\gamma p$ centre-of-mass energy is given by $W_{\gamma p}^2\simeq 2\omega\sqrt{s}$. The absorptive corrections due to spectator interactions between the two hadrons are represented by the factor $S_{\text{gap}}$. We will comment on the effect of absorption in  the next section.

The photon-Pomeron interaction will be described within the light-cone dipole frame. In this representation the probing
projectile fluctuates into a
quark-antiquark pair with transverse separation
$\rr$ long after the interaction, which then
scatters off the hadron \cite{nik}. The  amplitude for vector meson  production off nucleons reads  as \cite{nik},
\begin{eqnarray}
\, {\cal A}\, (x,Q^2,\Delta)  = \sum_{h, \bar{h}}
\int dz\, d^2\rr \,\Psi^\gamma_{h, \bar{h}}\,{\cal{A}}_{q\bar{q}}(x,r,\Delta) \, \Psi^{V*}_{h, \bar{h}} \, ,
\label{sigmatotp}
\end{eqnarray}
where $\Psi^{\gamma}_{h, \bar{h}}(z,\,\rr,Q^2)$ and $\Psi^{V}_{h,
  \bar{h}}(z,\,\rr)$ are the light-cone wavefunction  of the photon  and of the  vector meson, respectively. The
   quark and antiquark helicities are labeled by $h$ and $\bar{h}$, variable $\rr$ defines the relative transverse
separation of the pair (dipole), $z$ $(1-z)$ is the
longitudinal momentum fractions of the quark (antiquark). The quantity  $\Delta$ denotes the transverse momentum lost by the outgoing proton ($t = - \Delta^2$) and $x$ is the Bjorken variable. Moreover, ${\cal{A}}_{q\bar{q}}$ is the elementary amplitude for the scattering of a dipole of size $\rr$ on the target. It is related to the dipole cross section at forward limit, $\sigma_{dip}(x,r)=\mathrm{Im} {\cal{A}}_{q\bar{q}}(x,r,\Delta=0)$. Assuming the reasonable approximation of the $t$-dependence for the elementary cross section to be ${\cal{A}}_{q\bar{q}} \propto \exp \left(-B_V|t|/2 \right)$, the total cross section for exclusive production of charmonia off a nucleon target is given by,
\begin{eqnarray}
\sigma_{\gamma^* p\rightarrow Vp}(s,Q^2)  = \frac{1}{16\pi B_V} \left|{\cal A}\, (x,Q^2,\Delta=0)\right|^2,
\label{sigmatot}
\end{eqnarray} 
where $B_V$ is the diffractive slope parameter in the reaction $\gamma^*p\rightarrow V p$. Here, we consider the energy dependence of the slope using the Regge motivated expression $B_V(W_{\gamma p})=b_{el}^V + 2\alpha^{\prime}\log \left( \frac{W_{\gamma p}}{W_0}\right)^2$ with $\alpha^{\prime}=0.25$ GeV$^{-2}$ and $W_0=95$ GeV.  It is used the measured slopes \cite{H1psi2} for $\psi(1S)$ and $\psi(2S)$ at $W_{\gamma p}=90$ GeV, i.e. $b_{el}^{\psi(1S)}= 4.99\pm 0.41$ GeV$^{-2}$ and $b_{el}^{\psi(2S)}= 4.31\pm 0.73$ GeV$^{-2}$, respectively. 
In our numerical evaluations, the corrections related to skewedness effect and real part of amplitude are properly taken into account \cite{GM}. For the charm quark mass, we will use the value $m_c=1.4$ GeV.

The photon wavefunctions appearing in Eq. (\ref{sigmatotp}) are well known \cite{nik}. On the other hand, for the meson wavefunction we consider the boosted gaussian wavefunction:
\begin{eqnarray}
\psi_{\lambda, h\bar{h}}^{nS}  & = & \sqrt{\frac{N_c}{4\pi}} \frac{\sqrt{2}}{z(1-z)} \Biggl \{ \delta_{h,\bar{h}} \delta_{\lambda,2h} m_c +  i(2h)\delta_{h,-\bar{h}} e^{i\lambda\phi} \nonumber \\
&\times &\left[ (1-z)\delta_{\lambda,-2h} + z\delta_{\lambda,2h}   \right]\partial_r \Biggr \} \,\phi_{nS}(z,r) ,
\end{eqnarray}
where $\phi(z,r)$ in the mixed $(r,z)$ representation is obtained by boosting a Schr\"{o}dinger gaussian wavefunction in momentum representation, $\Psi(z,{\bf{k}})$. In this case, one obtains the following expression for the $1S$ state \cite{Sandapenpsi}: 
%------------------------------------------------
\begin{eqnarray}
\phi_{1S}(r,z) &=& N_T^{(1S)}\Biggl \{ 4z(1-z) \sqrt {2\pi R_{1S}^{2}}\exp\left[-{m_{q}^{2}R_{1S}^{2} \over 8z(1-z)}\right] \nonumber \\
& \times & \exp\left[-{2z(1-z)r^{2} \over R_{1S}^{2}}\right]
\exp\left[{m_{q}^{2}R_{1S}^{2} \over 2}\right] \Biggr \},
\label{eq:wf1S}
\end{eqnarray}
%------------------------------------------------
where for the $1S$ ground state vector mesons one determines the parameters
$R_{1S}^{2}$ and $N_T$ by considering the normalization property of wavefunctions and the predicted decay widths.

The radial wavefunction of the $\psi(2S)$ is obtained by the following modification of the $1S$ state \cite{Nemchik:1996pp}:
%------------------------------------------------
\begin{eqnarray}
& & \phi_{2S}(r,z) = N_T^{(2S)}\Biggl \{ 4z(1-z) \sqrt {2\pi R_{2S}^{2}}
\exp\left[-{m_{q}^{2}R_{2S}^{2} \over 8z(1-z)}\right] \nonumber \\
& \times & \exp\left[-{2z(1-z)r^{2} \over R_{2S}^{2}}\right]  
\exp\left[{m_{q}^{2}R_{2S}^{2} \over 2}\right] \nonumber \\
 &\times &  \left[ 1-\alpha \left( 1 + m_{q}^{2}R_{2S}^{2} -
{m_{q}^{2}R_{2S}^{2} \over 4z(1-z)}
+ {4z(1-z) \over R_{2S}^{2}}r^{2} \right)\right] \Biggr \}
 \nonumber\\
\label{eq:wf2S}
\end{eqnarray}
%------------------------------------------------
 with a new parameter $\alpha$. Now, the two parameters $\alpha$ and $R_{2S}$ are constrained from the
orthogonality conditions for the meson wavefunction. The choice of the meson wavefunction is one of main sources of theoretical uncertainty, introducing a typical 12 -- 13 \% error in theoretical prediction given a specific model for the dipole cross section.

Finally, here we will  consider the  phenomenological
saturation model proposed in Ref. \cite{IIM} (hereafter CGC model) which encodes the
main properties of the saturation approaches, with the dipole cross section  parameterized as follows
\begin{eqnarray}
\sigma_{dip}\,(x,\rr) =\sigma_0\,\left\{ \begin{array}{ll} 
{\mathcal N}_0 \left(\frac{\bar{\tau}^2}{4}\right)^{\gamma_{\mathrm{eff}}\,(x,\,r)}\,, & \mbox{for $\bar{\tau} \le 2$}\,, \nonumber \\
 1 - \exp \left[ -a\,\ln^2\,(b\,\bar{\tau}) \right]\,,  & \mbox{for $\bar{\tau}  > 2$}\,, 
\end{array} \right.
\label{CGCfit}
\end{eqnarray}
where $\bar{\tau}=\rr Q_{\mathrm{sat}}(x)$ and the expression for $\bar{\tau} > 2$  (saturation region)   has the correct functional
form, as obtained  from the theory of the Color Glass Condensate (CGC) \cite{hdqcd}. For the color transparency region near saturation border ($\bar{\tau} \le 2$), the behavior is driven by the effective anomalous dimension $\gamma_{\mathrm{eff}}\, (x,\,r)= \gamma_{\mathrm{sat}} + \frac{\ln (2/\tilde{\tau})}{\kappa \,\lambda \,y}$, where $\gamma_{\mathrm{sat}}=0.63$ is the LO BFKL anomalous dimension at saturation limit. 

\begin{figure}[t]
\includegraphics[scale=0.45]{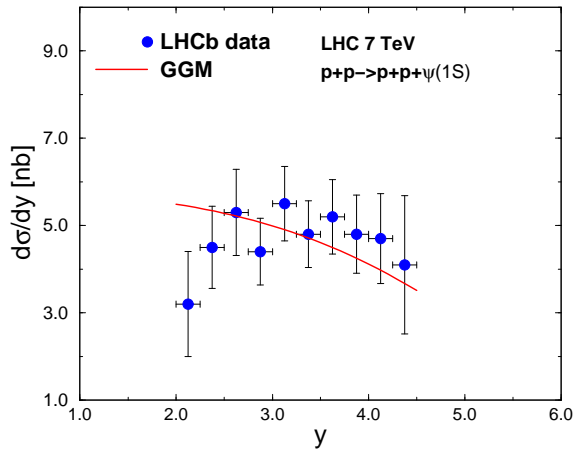}
\caption{(Color online) \it The rapidity distribution at forward region of exclusive $\psi(1S)$ meson production at $\sqrt{s}=7$ TeV in proton-proton collisions at the LHC. The theoretical prediction is labeled by solid curve (see text). Data from LHCb Collaboration \cite{LHCb}.}
\label{fig:1}
\end{figure}

\section{Results and discussions}
\label{resultados}

\begin{figure}[t]
\includegraphics[scale=0.45]{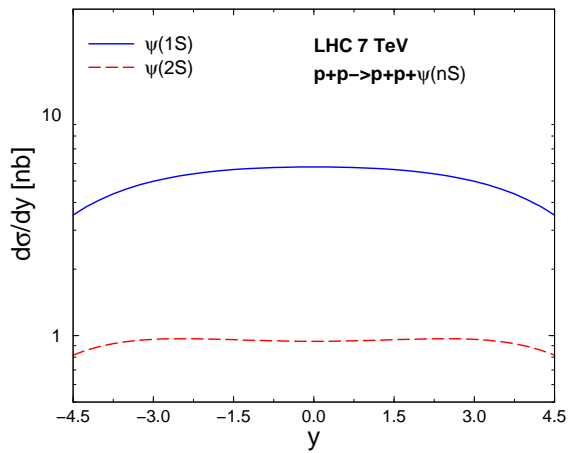}
\caption{(Color online) \it The theoretical predictions for the rapidity distribution of exclusive $J/\psi$ (solid curve) and $\psi(2S)$ (dashed curve) meson production at $\sqrt{s}=7$ TeV in proton-proton collisions at the LHC.}
\label{fig:2}
\end{figure}

First, we compare the present theoretical approach to the data for the $\psi (1S)$ state measured by LHCb Collaboration at the energy of 7 TeV in proton-proton collisions at the forward region $2.0<\eta_{\pm} <4.5$ \cite{LHCb}. Hereafter, we will assume for the absorption factor the average value $S_{\text{gap}}^2=0.8$, despite it depends on rapidity as shown in Ref. \cite{SS}. The absorptive corrections considering the elastic rescattering have been computed for $pp$ collisions in \cite{SS} and have a value of $S_{\text{gap}}^2(y=0)=0.85$ and  $S_{\text{gap}}^2 (y=3)=0.75$, respectively. In Fig. \ref{fig:1} the numerical calculations (labeled GGM and represented by the solid curve)  are shown for the rapidity distribution for $\psi(1S)$ state within the color dipole formalism, Eqs. (\ref{dsigdy}) and (\ref{sigmatot}). The relative normalization and overall behavior on rapidity is quite well reproduced in the forward regime. In Fig. \ref{fig:2} the complete rapidity distribution, including mid-rapidity and backward region, is presented for $J/\psi$ and $\psi (2S)$ states (solid and dashed curves, respectively). The $J/\psi$ cross section at central rapidity is $\frac{d\sigma}{dy}(y=0)= 5.8$ nb. We obtain $\sigma (pp\rightarrow p+J/\psi+p) \times \mathrm{Br}(J/\psi \rightarrow \mu^+\mu^-)=698$ pb for the meson with a rapidity between 2 and 4.5. After correcting this result by the acceptance factor in order to convert the prediction in terms of muon pseudorapidities we get $\sigma_{pp\rightarrow J/\psi(\rightarrow \mu^+\mu^-)}(2.0<\eta_{\mu^\pm}<4.5) = 298$ pb. This is in good agreement to the experimental result $\sigma_{pp\rightarrow J/\psi(\rightarrow \mu^+\mu^-)}(2.0<\eta_{\mu^\pm}<4.5) = 307\pm 42$ pb \cite{LHCb} (summing errors in quadrature).

Now, we analyze the exclusive production of the radially excited $\psi(2S)$ mesons. In  Fig. \ref{fig:2} we show the rapidity distribution for that meson state (dashed line), which gives at central rapidity a cross section $\frac{d\sigma}{dy}(y=0)= 0.94$ nb. It is obtained $\sigma (pp\rightarrow p+\psi (2S)+p) \times \mathrm{Br}(\psi (2S) \rightarrow \mu^+\mu^-)=18$ pb for rapidities $2.0<y<4.5$. Accordingly, we now predict $\sigma_{pp\rightarrow \psi(2S)(\rightarrow \mu^+\mu^-)}(2.0<\eta_{\mu^\pm}<4.5) = 7.7$ pb compared to  $\sigma_{pp\rightarrow \psi(2S)(\rightarrow \mu^+\mu^-)}(2.0<\eta_{\mu^\pm}<4.5) = 7.8\pm 1.6$ pb measured by LHCb \cite{LHCb}. At mid-rapidity we obtain the ratio $[\psi (2S)/\psi (1S)]_{y=0}=0.16$ and taking the integrated cross section for $2.0<y<4.5$ we have $[\psi (2S)/\psi (1S)]_{2<y<4.5}=0.18$. The latter value of the ratio is strongly consistent to the LHCb determination $[\psi (2S)/\psi (1S)] (2.0<\eta_{\mu^\pm}<4.5)=0.19\pm 0.04$.

It is timely to compare our results to the similar theoretical approaches in literature. The values obtained for the integrated cross sections for the exclusive $J/\psi$ production are consistent with calculations using the color dipole formalism \cite{GM,MW} and with the prediction from Starlight \cite{STARLIGHT} and SuperChic \cite{SUPERCHIC} generators as well. In the case of $\psi (2S)$ state our prediction is in agreement to the Starlight generator result, which gives $\sigma^{\mathrm{STARLIGHT}}_{pp\rightarrow \psi(2S)(\rightarrow \mu^+\mu^-)}= 6.1$ pb. On the other hand, our results are a factor about 2 lower than the values appearing in Ref. \cite{SS} that considers the $k_{\perp}$-factorization approach. 

The complexity of the phenomenological model considered here could turn out its connection to the QCD dynamics not so clear. Therefore, some comments are in order at this point. The main dependence being probed in the exclusive vector meson production in $pp$ collisions is the energy behavior of the photoproduction cross section (give that the photon flux is well known). The LHCb data cover one order of magnitude on photon-proton centre-of-mass energy above the typical DESY-HERA regime. Such an extrapolation is completely driven by the QCD dynamics at small-x  and embeded in the dipole cross section in our analysis. This is directly translated into the rapidity dependence of $pp$ cross section. The other inputs like the paramaters in wavefunctions, real part and skwedness corrections and absorption effects only account for the overall normalization. It's worth mentioning that our predictions are parameter free. The inputs in the meson wavefunction are determined from its normalization condition. The phenomenological parameters in dipole cross section were determined by a fit to DESY-HERA data for proton structure function $F_2$ at small-x \cite{IIM} and already tested against exclusive  processes at DESY-HERA energies at a number of contributions \cite{Sandapenpsi,MW}.

Finally, we perform predictions for the next LHC runs in proton-proton mode. We have found $\frac{d\sigma_{J/\psi}}{dy} = 6.2$ nb and  7.9 nb for central rapidities at energies of 8 and 14 TeV, respectively. For the $\psi (2S)$ state the extrapolation gives $\frac{d\sigma_{\psi(2S)}}{dy} = 1.0$ nb and 1.4 nb for the same energies at central rapidity. We have checked that our values are lower than those predicted by the $k_{\perp}$-factorization approach \cite{SS}, which gives $\frac{d\sigma_{\psi(1S)}}{dy} \simeq 11$ nb  and $\frac{d\sigma_{\psi(2S)}}{dy} \simeq 2 $ nb for energy of 14 TeV and central rapidity.

\section{Summary}
\label{conc}

We have investigated the exclusive $J/\psi$ and  radially excited $\psi (2S)$ photoproduction off nucleons in proton-proton collisions at the LHC. The theoretical framework considered in the analysis is the light-cone dipole formalism and predictions are obtained for centre-of-mass energies 7, 8 and  14 TeV.  It was found that the coherent exclusive photoproduction of $\psi(2S)$  off nuclei has an upper bound of order 0.71 mb at $y=0$ down to 0.10 mb for backward/forward rapidities $y=\pm 3$. The incoherent contribution was also computed and it is a factor 0.2 below the coherent one.  Comparison has been done to the recent  LHCb Collaboration data for the exclusive $\psi(1S)$ and $\psi (2S)$ production at 7 TeV.  The experimental values are fairly described by the present calculation, which gives $\sigma [pp\rightarrow J/\psi(\rightarrow \mu^+\mu^-)] = 298$ pb, $\sigma [pp\rightarrow \psi(2S)(\rightarrow \mu^+\mu^-)]= 7.7$ pb and  $R\left(\frac{\psi (2S)}{\psi (1S)}\right)=0.18$ in the pseudo-rapidity range $2.0<\eta_{\mu^\pm}<4.5$.

\begin{acknowledgments}
This work was  partially financed by the Brazilian funding
agencies CNPq and FAPERGS and by the French-Brazilian scientific cooperation project CAPES-COFECUB 744/12. 
\end{acknowledgments}

\end{document}